\documentclass[11pt, a4paper,submission, Proceedings]{SciPost}
\pdfoutput=1

\hypersetup{
    colorlinks,
    linkcolor={red!50!black},
    citecolor={blue!50!black},
    urlcolor={blue!80!black}
}

\usepackage[bitstream-charter]{mathdesign}
\DeclareSymbolFont{usualmathcal}{OMS}{cmsy}{m}{n}
\DeclareSymbolFontAlphabet{\mathcal}{usualmathcal}

\usepackage{wrapfig}
\usepackage{lipsum}
\usepackage{cleveref}
\Crefname{figure}{Fig.}{Figs.}
\Crefname{equation}{Eq.}{Eqs.}

\begin{document}

\begin{center}{\Large \textbf{
Cosmic Ray Measurements with IceCube and IceTop\\
}}\end{center}

% TODO: write the author list here. Use initials + surname format.
% Separate subsequent authors by a comma, omit comma at the end of the list.
% Mark the corresponding author with a superscript *.
\begin{center}
Dennis Soldin\textsuperscript{1,2*} for the IceCube Collaboration
\end{center}

\begin{center}
{\bf 1} Karlsruhe Institute of Technology, Institute of Experimental Particle Physics, 76021 Karlsruhe, Germany
\\
{\bf 2} Bartol Research Institute and Dept. of Physics and Astronomy, University of Delaware, Newark, DE 19716, USA
\\
* soldin@kit.edu
\end{center}

%\vspace{-1.5em}

\begin{center}
\today
\end{center}

% For convenience during refereeing (optional),
% you can turn on line numbers by uncommenting the next line:
%\linenumbers
%\vspace{-2em}

\definecolor{palegray}{gray}{0.95}
\begin{center}
\colorbox{palegray}{
  \begin{tabular}{rr}
  \begin{minipage}{0.1\textwidth}
    \includegraphics[width=30mm]{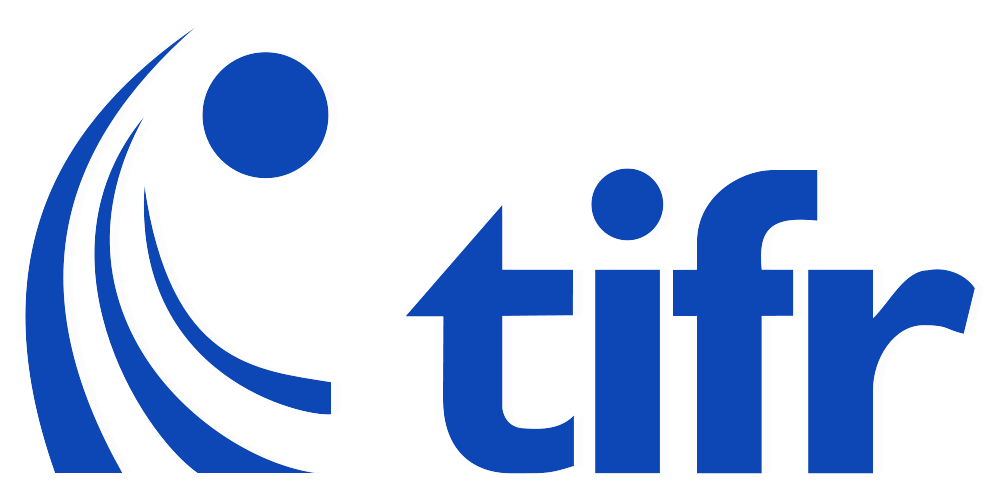}
  \end{minipage}
  &
  \begin{minipage}{0.85\textwidth}
    \begin{center}
    {\it 21st International Symposium on Very High Energy Cosmic Ray Interactions (ISVHECRI 2022)}\\
    {\it Online, 23-27 May 2022} \\
    \doi{10.21468/SciPostPhysProc.?}\\
    \end{center}
  \end{minipage}
\end{tabular}
}
\end{center}

\vspace{-2em}

\section*{Abstract}
{\bf
% TODO: write your abstract here.
IceCube is a cubic-kilometer Cherenkov detector in the deep ice at the geographic South Pole. The dominant event yield in the deep ice detector consists of penetrating atmospheric muons with energies above approximately 300\,GeV, produced in cosmic ray air showers. In addition, the surface array, IceTop, measures the electromagnetic component and GeV muons of air showers. Hence, IceCube and IceTop yield unique opportunities to study cosmic rays with unprecedented statistics in great detail.

We will present recent results of comic ray measurements from IceCube and IceTop. In this overview, we will highlight measurements of the energy spectrum of cosmic rays from 250\,TeV up to the EeV range and their mass composition above 3\,PeV. We will also report recent results from measurements of the muon content in air showers and discuss their consistency with predictions from current hadronic interaction models.

%an analysis of the density of GeV muons observed in IceTop and discuss their consistency with predictions from current hadronic interaction models.
}

\vspace{10pt}
\noindent\rule{\textwidth}{1pt}
%\vspace{-2em}
\tableofcontents\thispagestyle{empty}
%\vspace{-1em}
\noindent\rule{\textwidth}{1pt}
%\vspace{10pt}

\section{Introduction}
\label{sec:intro}

High-energy cosmic rays enter the Earth's atmosphere and interact with air molecules at energies equivalent to those at current collider experiments, such as the Large Hadron Collider (LHC), and higher. Their energy spectrum has been measured with high precision over $11$ orders of magnitude, up to energies of a few $100\,\mathrm{EeV}$. However, the sources of cosmic rays are still unknown, their acceleration mechanisms and mass composition are uncertain, and several features in the energy spectrum are not well understood~\cite{Kampert:2012mx}. Large uncertainties in the understanding of cosmic rays remain because measurements at energies above $100\,\mathrm{TeV}$ present significant experimental challenges. Cosmic ray interactions produce large particle cascades  in the atmosphere, extensive air showers (EASs), which are measured with large detector arrays at the ground. The properties of the initial cosmic rays, such as their energy and mass, are determined indirectly and thus the interpretation of these measurements strongly relies on theoretical models of the EAS development in the atmosphere. In recent years, large discrepancies have been observed between experimental data and model predictions which provide stringent limitations for the interpretations~\cite{Albrecht:2021yla}. In this report, we will present an overview of recent results from cosmic ray measurements by the IceCube Neutrino Observatory.

\section{The IceCube Neutrino Observatory}
\label{sec:icecube}

\begin{wrapfigure}{r}{0.42\textwidth}
\vspace{-3.5em}
\includegraphics[width=.43\textwidth]{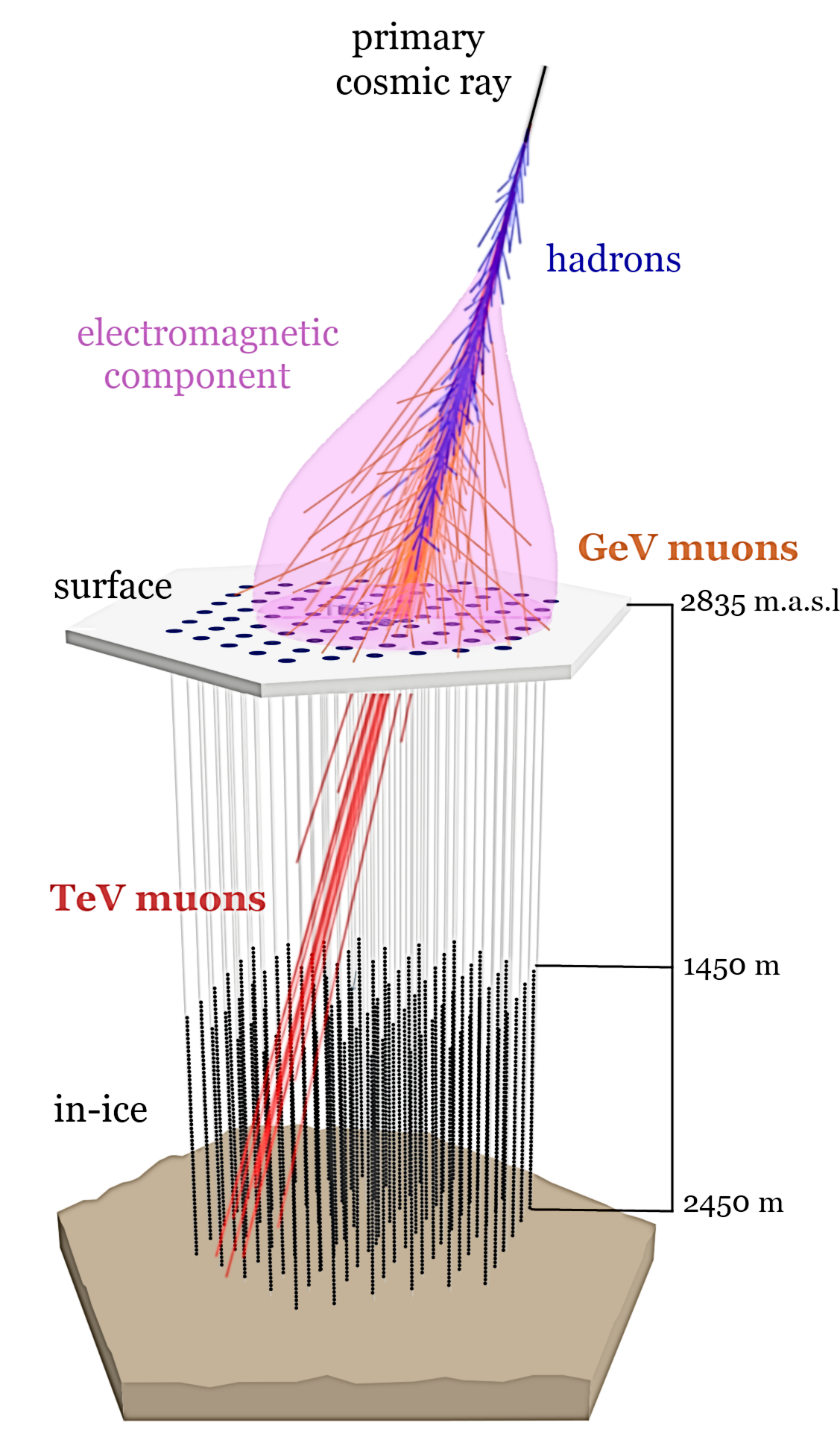}
\vspace{-2.3em}

\caption{Schematic picture of the detection of cosmic rays with IceCube.}
\label{fig:IceCube}
\vspace{-1em}
\end{wrapfigure}

The IceCube Neutrino Observatory (IceCube)~\cite{Aartsen:2016nxy} is a large-scale particle detector, located at the geographic South Pole. IceCube consists of a cubic-kilometer detector in the deep Antarctic ice, accompanied by a square-kilometer detector array at the surface, IceTop~\cite{IceCube:2012nn}, as shown in~\Cref{fig:IceCube}.
The in-ice detector of IceCube is located at depths between about $1.5\,\mathrm{km}$ and $2.5\,\mathrm{km}$, comprising $86$ strings with more than $5100$ digital optical modules (DOMs). The strings are deployed in an hexagonal array with an average spacing of $125\,\mathrm{m}$. The average trigger rate of about $2.15\,\mathrm{kHz}$ is mainly caused by high-energy atmospheric muons with typical energies around several $100\,\mathrm{GeV}$ which penetrate the ice and generate Cherenkov light that is measured by the DOMs. IceTop is located at an altitude of about $2.8\,\mathrm{km}$ above sea level, corresponding to an atmospheric depth of $\sim 690\,\mathrm{g/cm}^2$. IceTop comprises $81$ stations with each station consisting of two cylindrical ice-Cherenkov tanks and housing two DOMs each, deployed approximately consistent with the location of the IceCube strings. An infill area in the center of the surface detector has a denser spacing of $<50\,\mathrm{m}$ which is used to improve the sensitivity of air shower detection at low energies. The IceTop tanks are able to measure the electromagnetic component of the air shower and low-energy muons. The complementary information from both detectors yields unique opportunities to perform a variety of measurements of primary cosmic rays with energies from a few $100\,\mathrm{TeV}$ up to the EeV range, as described in the following.

\section{Cosmic Ray Flux}
\label{sec:spectrum_composition}

\subsection{Energy Spectrum}
\label{sec:spectrum}

The properties of cosmic rays can be determined based on the EAS signals measured by the IceTop tanks. The tank signals are calibrated to account for the specific tank response and after calibration they are expressed in units of expected vertical equivalent muons (VEM)~\cite{IceCube:2012nn}. For further analysis, various event cleanings are applied, as described in Ref.~\cite{PhysRevD.88.042004}. An air shower reconstruction is then applied to the data to determine the position and direction of the shower axis and the signal distribution is fit by a lateral distribution function (LDF) of the form
\begin{equation}
S(r)=S_{125}\cdot \left(\frac{r}{125\,\mathrm{m}}\right)^{-\beta-\kappa\cdot \log_{10}(r/125\,\mathrm{m})}\,.
\label{eq:LDF}
\end{equation}
This function describes the signal distribution of the event, $S(r)$, in units of VEM, as a function of the lateral distance, $r$. The parameters $\beta$ and $S_{125}$ are free during the fitting procedure and they measure the steepness and the signal strength at a reference distance of $125\,\mathrm{m}$, respectively. The constant parameter $\kappa=0.303$ has been obtained from simulations and it describes the curvature of the log-parabola. The timing distribution of the signals is fit simultaneously by a paraboloid with a Gaussian nose, as defined in Ref.~\cite{PhysRevD.88.042004}. The best fit parameters $\beta$ and $S_{125}$ are obtained using a three-step maximum-likelihood method where the charges and timing of the measured signals are compared to the expected charge and timing distributions.

Due to the environmental conditions at the South Pole, snow accumulates on top of the IceTop tanks with the depth depending on the location and the time of data taking. The depth of the snow varies between tens of centimeters and a few meters. In order to account for the snow accumulation during the EAS reconstruction an exponential reduction of the tank signals is applied to the expected signal strength, $S(r)$, as described in Ref.~\cite{IceCube:2012nn}. The uncertainties due to this procedure are included in the detector systematics.

\begin{figure}[!b]
\centering
\vspace{-1em}

\includegraphics[width=0.85\textwidth]{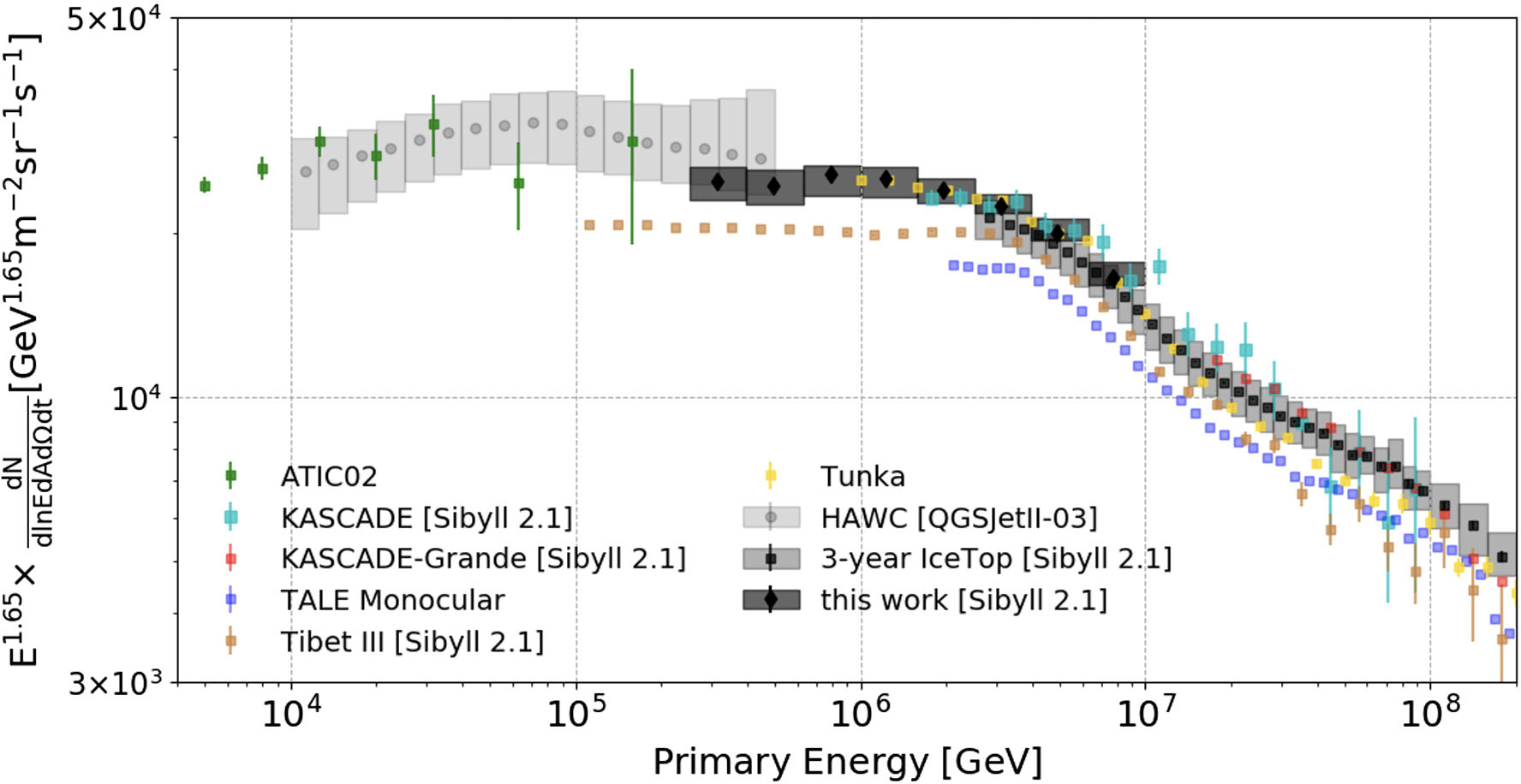}
\vspace{-0.6em}

\caption{Cosmic ray energy spectrum measured with IceTop in comparison to results from other experiments. Figure is taken from Ref.~\cite{IceCube:2020yct}.}
\label{fig:CRSpectrum}
\end{figure}

After EAS reconstruction, additional quality cuts are applied (see Ref.~\cite{PhysRevD.88.042004}) and the zenith angles of surviving events is restricted to angles below $\sim 37^\circ$ for which the reconstruction algorithm is optimized. The best fit $S_{125}$ is then used as an energy proxy which is converted into the EAS energy, $E_0$. This is done for different zenith angles, $\theta$, using CORSIKA simulations~\cite{corsika} with Sibyll~2.1~\cite{Ahn:2009wx} as the hadronic interaction model and assuming an H4a cosmic ray flux model~\cite{Gaisser:2011cc}. Using this technique, the resulting energy resolution is below $0.1$ in $\log_{10}(E_0/\mathrm{GeV})$ for all energies considered in this analysis.

Based on this technique, the energy spectrum of cosmic rays was determined using IceTop data from June 2010 through May 2013~\cite{IceCube:2019}. The resulting energy spectrum of cosmic rays above $2.5\,\mathrm{PeV}$ is shown in \Cref{fig:CRSpectrum} (\emph{``3-year IceTop''}). Systematic uncertainties include uncertainties due to the VEM calibration and snow accumulation and they are shown as a grey band. Also shown are results from other experiments for comparison (for details see Ref.~\cite{IceCube:2020yct}).

In order to extend the measurement of the spectrum towards lower energies, a dedicated event selection is needed which uses the denser infill area of IceTop~\cite{IceCube:2020yct}. However, the EAS reconstruction technique described above is not feasible for events with only a few detector stations hit. Instead, the reconstruction of low-energy events is based on an iterative random forest regression technique, as described in Ref.~\cite{IceCube:2020yct}. The random forest
is trained using $50\%$ of the simulated events and the other $50\%$ are used for testing and performance optimization which is done using a cross-validation grid search. The CORSIKA simulations used in this analysis use Sibyll~2.1 as the hadronic interaction model and an H4a cosmic ray flux is assumed. To account for efficiency effects in this analysis, an efficiency correction is applied to the data. In addition, a Bayesian iterative unfolding is used to account for potential bin migration effects. Because the random forest is trained with simulations using certain model assumptions, systematic biases due to the cosmic ray flux, the atmospheric model, the efficiency correction, and the unfolding procedure are included as systematic uncertainties. The resulting energy resolution is better than $0.2$ in $\log_{10}(E_0/\mathrm{GeV})$ over the entire energy range considered.

The low-energy spectrum was determined from IceTop data taken between May 2016 and May 2017. The resulting energy spectrum is shown in \Cref{fig:CRSpectrum} (\emph{``this work''}) for energies between $250\,\mathrm{TeV}$ and $10\,\mathrm{PeV}$ in comparison to the high-energy spectrum. Both results agree in the overlap region within systematic uncertainties.

\subsection{Mass Composition}
\label{sec:composition}

\begin{figure}[!b]
\vspace{-0.6 cm}

  \mbox{\hspace{-0.0 cm}
  \includegraphics[width=0.575\textwidth]{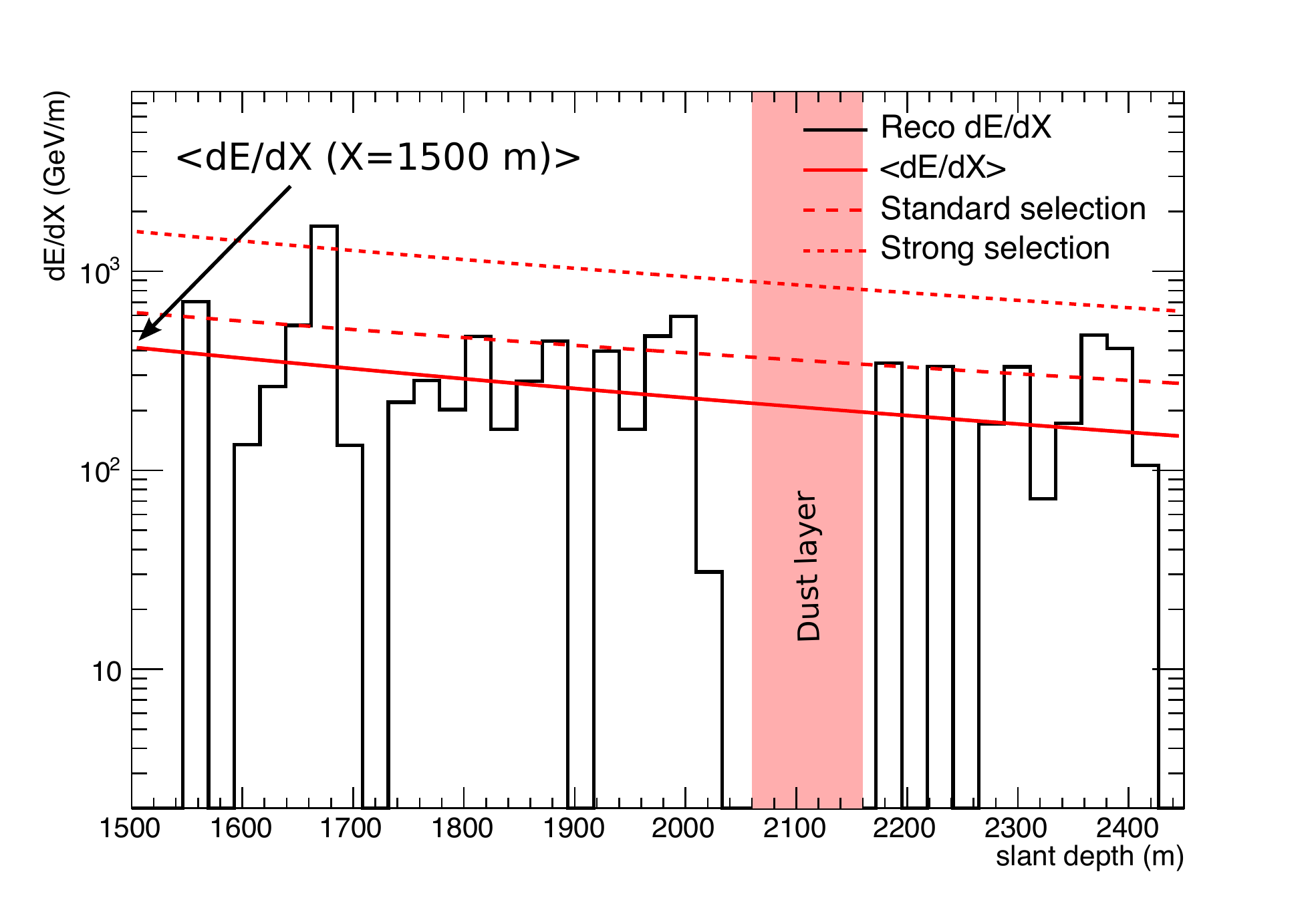}\hspace{-0.6cm}
  \includegraphics[width=0.4\textwidth]{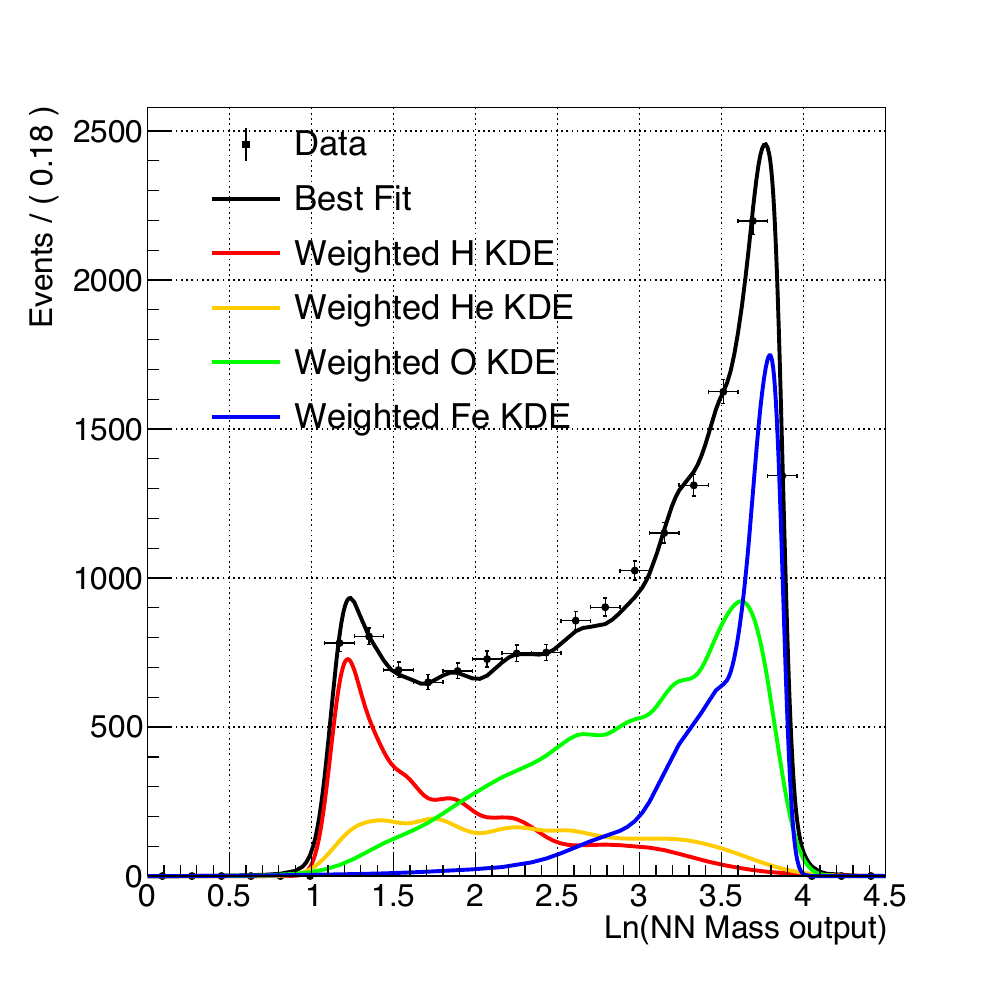}}
  \vspace{-.5em}
  
  \caption{\emph{Left:} Energy loss profile obtained using the reconstruction algorithm described in Ref.~\cite{IceCube:2013dkx}. In addition to the fit to the continuous energy loss (solid), two different thresholds used to count large stochastic losses are also shown (dashed/dotted). \emph{Right:} Mass output from the neural network for the energy bin $7.4\leq \log_{10}(E_0/\mathrm{GeV})\leq 7.5$. Colored lines show the PDF templates that are fit to the data in order to obtain the mass composition. Figures are taken from Ref.~\cite{IceCube:2019}.}
  \label{fig:CRMassTemplate}
  \vspace{-0.3cm}
  
\end{figure}

According to the Matthews-Heitler model \cite{Matthews:2005sd}, the number of muons produced in EASs, $N_\mu$, scales with the mass number, $A$, and the energy of the initial cosmic ray, $E_0$, as
\begin{equation}
N_\mu(E_0,A)\propto A\cdot\left(\frac{E_0}{A}\right)^\beta\, ,
\end{equation}
with $\beta\simeq 0.9$. Moreover, protons are more likely to produce high-energy muons than heavier nuclei which can have large energy deposits in a dense medium, like ice, due to radiative processes, such as Bremsstrahlung and pair-production~\cite{Aartsen:2015nss}. Thus, measurements of the muon content in EASs via their energy losses in the deep portion of IceCube, together with information from IceTop, can be used to estimate the mass spectrum of cosmic rays.

\begin{figure}[tb]
\centering
  \vspace{-0.5cm}
  
\includegraphics[width=.98\textwidth]{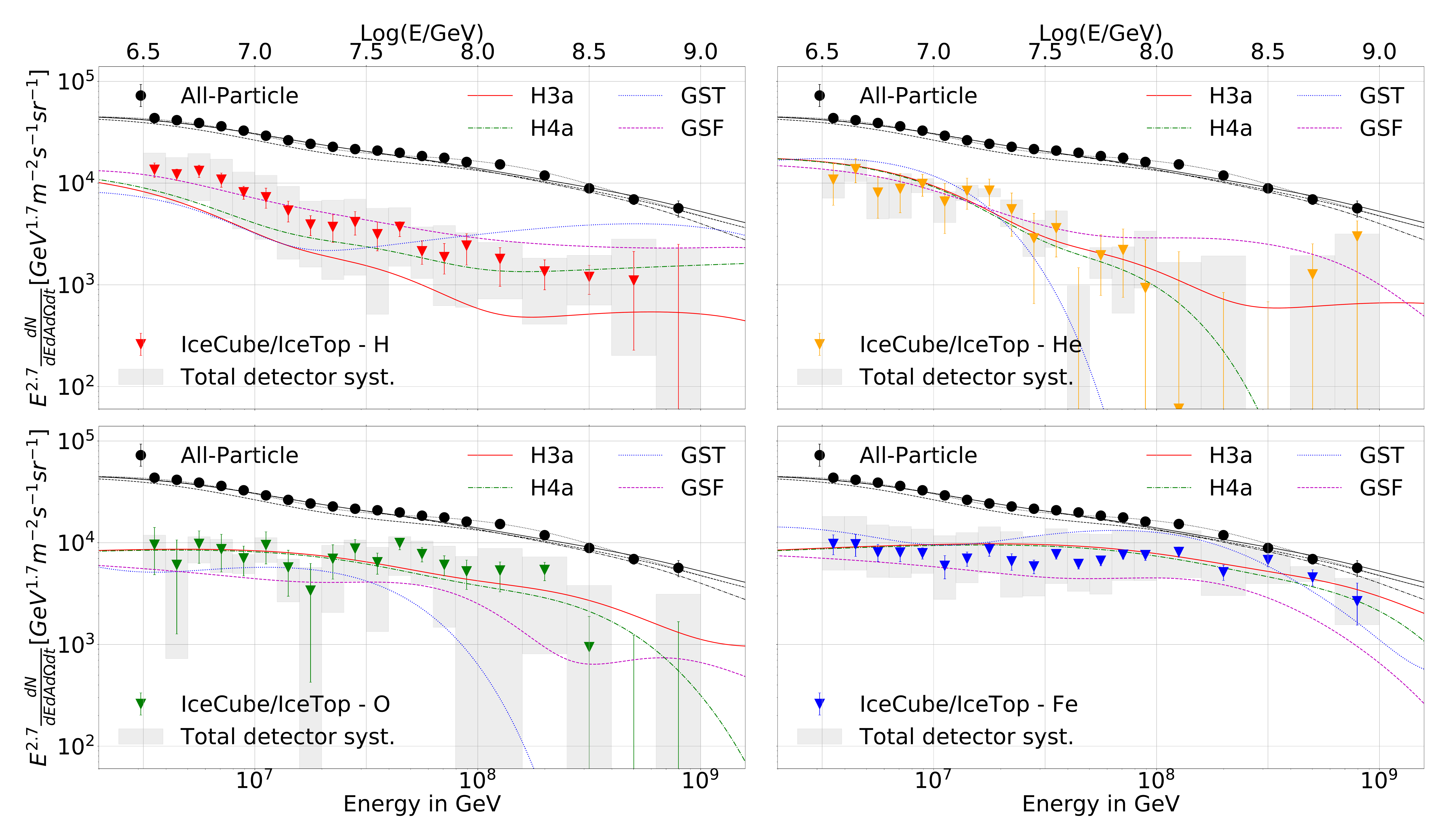}
  \vspace{-1em}
  
\caption{Cosmic ray mass spectra, obtained from IceCube data taken between June 2010 and May 2013, for four different mass groups. Also shown are the detector systematics (grey bands) and predictions  for the flux models H3a/H4a~\cite{Gaisser:2011cc}, GST~\cite{Stanev:2014mla}, and GSF~\cite{Dembinski:2015xtn}. Figure is taken from Ref.~\cite{IceCube:2019}.}
  \vspace{-0.3cm}
  
\label{fig:CRMass}
\end{figure}

The mass composition of cosmic rays is obtained using coincident events observed in both the deep ice detector of IceCube and IceTop. The same selection criteria are applied as for the analysis of the high-energy cosmic ray spectrum, described above. The deposited energy along the extrapolated shower axis in the ice, $dE/dX$, is derived based on the charge and timing information using a dedicated reconstruction algorithm~\cite{IceCube:2013dkx}. An example of a reconstructed energy loss profile of high-energy muons in the ice is shown in \Cref{fig:CRMassTemplate} (left). The reconstructed energy loss at a reference slant depth of $1500\;\mathrm{m}$ in the ice, as well as two measures of the number of high-energy stochastic losses along the reconstructed trajectory, as defined in Ref.~\cite{IceCube:2019}, are used as mass sensitive variables for further analysis. These observables, together with the energy proxy, $S_{125}$, and the zenith angle obtained from IceTop, are used as input for an artificial neural network (ANN)~\cite{Voss:2009rK} to determine the initial cosmic ray energy and mass. The ANN is trained using $50\%$ of CORSIKA simulations based on Sibyll~2.1 with equal fractions of H, He, O, an Fe primary cosmic rays. The other $50\%$ of the events in simulations is used to test and optimize the ANN performance, as described in Ref.~\cite{IceCube:2019}. The ANN output distributions are then converted into template probability density functions (PDFs) using an adaptive kernel density estimation (KDE) method~\cite{Cranmer:2000du}. These PDFs are used to fit the data in each energy bin separately. An example PDF fit for one energy bin is shown in \Cref{fig:CRMassTemplate} (right).

The resulting cosmic ray all-particle spectrum, obtained from data taken between June 2010 and May 2013, as well as individual mass spectra for the four mass groups used in this analysis, are shown in \Cref{fig:CRMass}~\cite{IceCube:2019}. Systematic uncertainties are shown as grey bands and they include effects of the snow accumulation, the absolute energy scale, and the light yield measured in the ice. Also shown for comparison are predictions for the cosmic ray flux models H3a/H4a~\cite{Gaisser:2011cc}, GST~\cite{Stanev:2014mla}, and GSF~\cite{Dembinski:2015xtn}. The all-particle spectrum agrees well with the spectrum discussed in \Cref{sec:spectrum} which is obtained using independent analysis methods.

\section{Muon Content in Air Showers}
\label{sec:muons}

\subsection{GeV Muon Density}
\label{sec:GeVmuons}

\begin{figure}[!b]
\centering
 \vspace{-0.3cm}
  
\mbox{\includegraphics[width=0.42\textwidth]{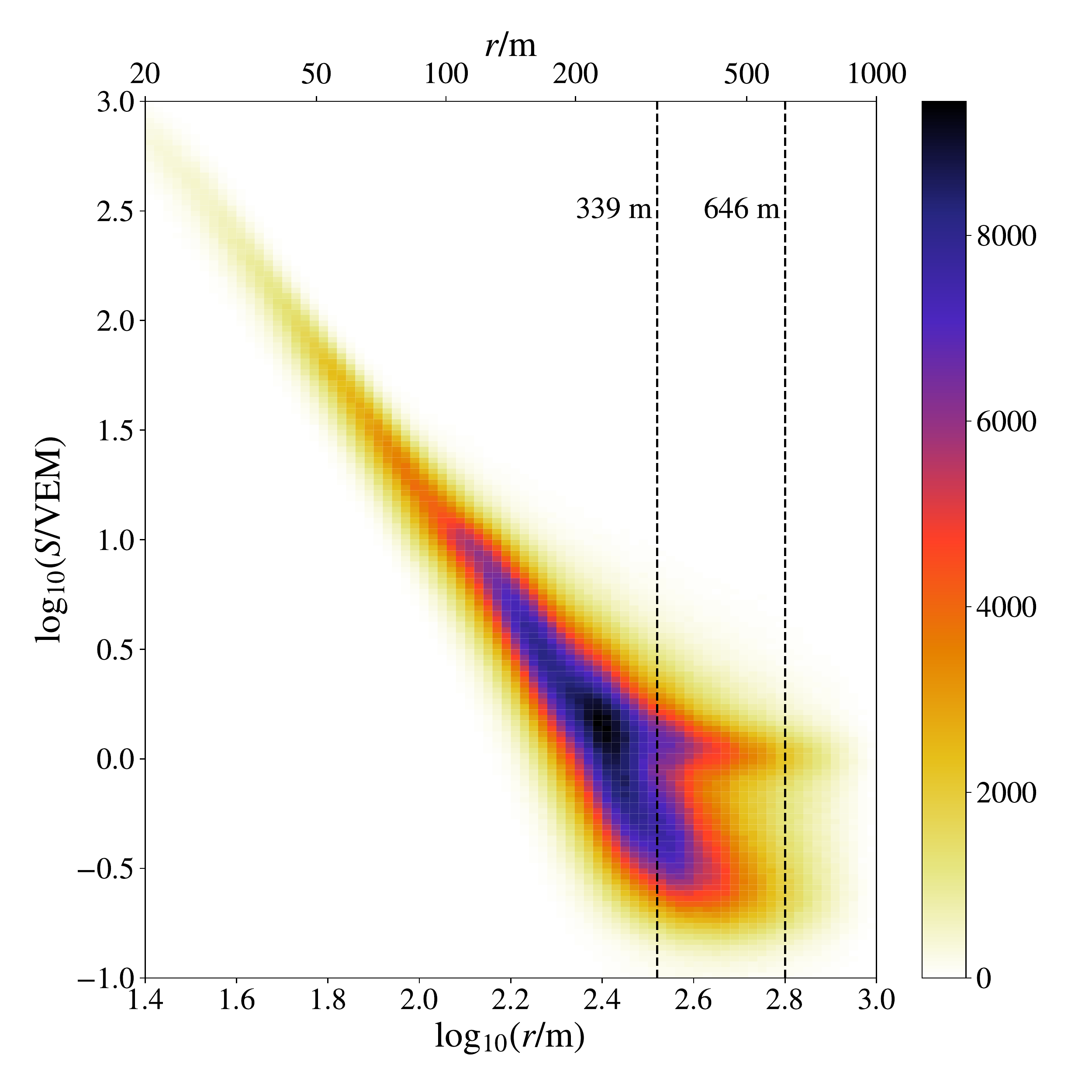}%\hspace{-1em}
\includegraphics[width=0.53\textwidth]{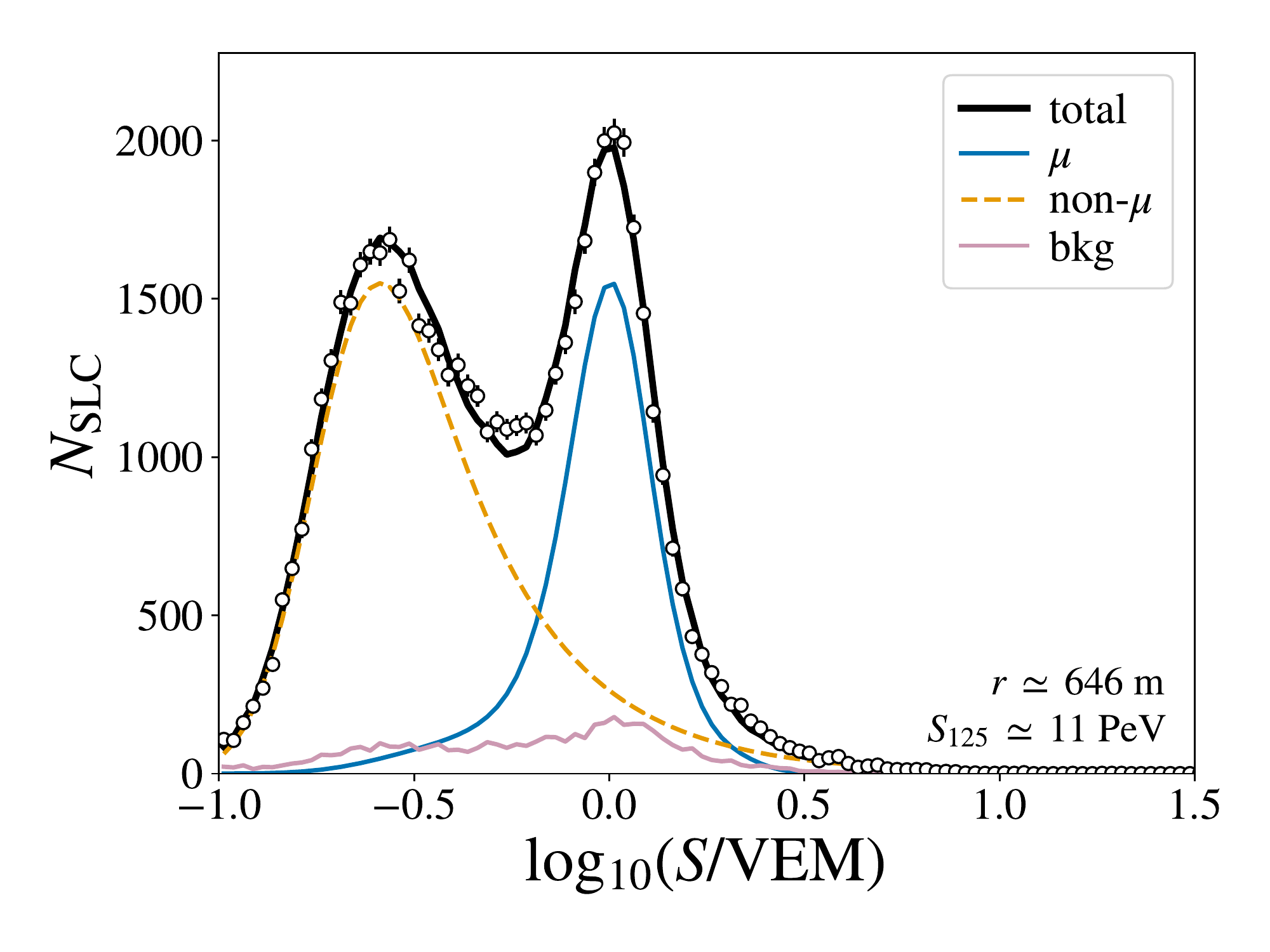}}
 \vspace{-1em}
 
\caption{\emph{Left:} IceTop signals as a function of the lateral distance, r, for near-vertical EASs ($\theta\leq 18^\circ$) with energies between $10\,\mathrm{PeV}$ and $12.5\,\mathrm{PeV}$. \emph{Right:} Signal distribution at a lateral distance of $646\,\mathrm{m}$ and fit to the semi-analytical signal model that accounts for an electromagnetic component, the muon component, as well as for uncorrelated background. Figures are taken from Ref.~\cite{IceCube:2022yap}.}
\label{fig:MuonThumb}
\end{figure}

%\begin{wrapfigure}{r}{0.5\textwidth}
%\vspace{-4em}
%\includegraphics[width=.5\textwidth]{plots/MuonDensity0}

%\includegraphics[width=.5\textwidth]{plots/MuonDensity1}

%\includegraphics[width=.5\textwidth]{plots/MuonDensity2}
%\vspace{-2.3em}

%\caption{Muon densities in terms of the z-values defined in \Cref{eq:zvalues}, compared to predictions from the hadronic interaction models Sibyll~2.1, EPOS-LHC, and QGSJet-II.04. Also shown are the expectations from the cosmic ray flux models H3a~\cite{Gaisser:2011cc}, GST~\cite{Stanev:2014mla}, and GSF~\cite{Dembinski:2015xtn}. Figures are taken from Ref.~\cite{IceCube:2022yap}.}
%\label{fig:RhoMu}
%\vspace{-1em}
%\end{wrapfigure}

IceTop is also sensitive to low-energy muons with typical energies of about $1\,\mathrm{GeV}$. While the bulk of EAS particles close to the shower axis is dominated by electrons and photons, muons become the dominant particles at large distances from the shower axis. \Cref{fig:MuonThumb} (left) shows the distribution of tank signals, $S(r)$, as a function of the distance from the reconstructed shower axis, $r$, at EAS energies between $10\,\mathrm{PeV}$ and $12.5\,\mathrm{PeV}$ for near-vertical showers ($\theta \leq 18^\circ$). The majority of signals follows the LDF defined in \Cref{eq:LDF}, however, at large distances a structure around $S(r)=1\,\mathrm{VEM}$ becomes visible. This population consists of tank signals produced by single muons with typical energies of $\sim 1\,\mathrm{GeV}$ and it is used to determine the low-energy muon content in EASs. This is done using vertical slices in \Cref{fig:MuonThumb} (left) at a fixed EAS energy, zenith angle, and lateral distance, based on the EAS reconstruction described in \Cref{sec:spectrum}. An example distribution at a lateral distance of $646\,\mathrm{m}$ is shown in  \Cref{fig:MuonThumb} (right) which is fit using a multi-component semi-analytical model that accounts for an electromagnetic component, the muon component, as well as for uncorrelated background signals, as described in Ref.~\cite{IceCube:2022yap}. During the reconstruction procedure the effect due to the snow accumulation is taken into account and multiple free parameters are fit. Most importantly for this analysis is the mean number of muons, $\langle N_\mu\rangle$, which is divided by the cross-sectional area of the IceTop tanks to determine the muon density, $\rho_\mu$. Systematic uncertainties due to snow accumulation, the absolute energy scale calibration, and the electromagnetic model in the likelihood fit are considered as discussed in detail in Ref.~\cite{IceCube:2022yap}. To account for small differences in the reconstructed muon density between simulation and data, a correction is applied and the corresponding uncertainties are included in the systematics.

\begin{figure}[tb]
\centering
\vspace{-0em}
\mbox{\hspace{-3.em}\includegraphics[width=0.411\textwidth]{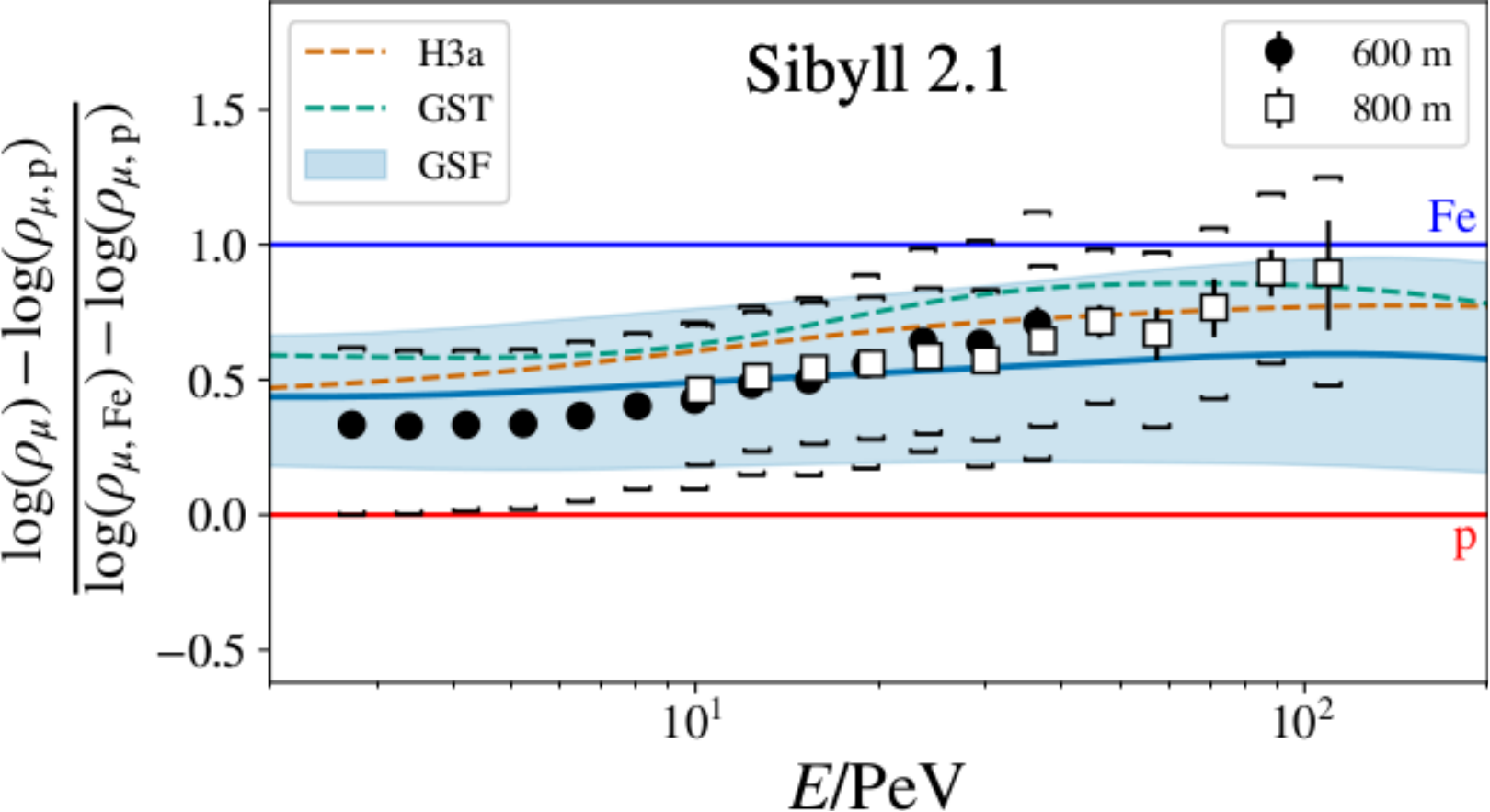}\hspace{-1mm}
\includegraphics[width=0.34\textwidth]{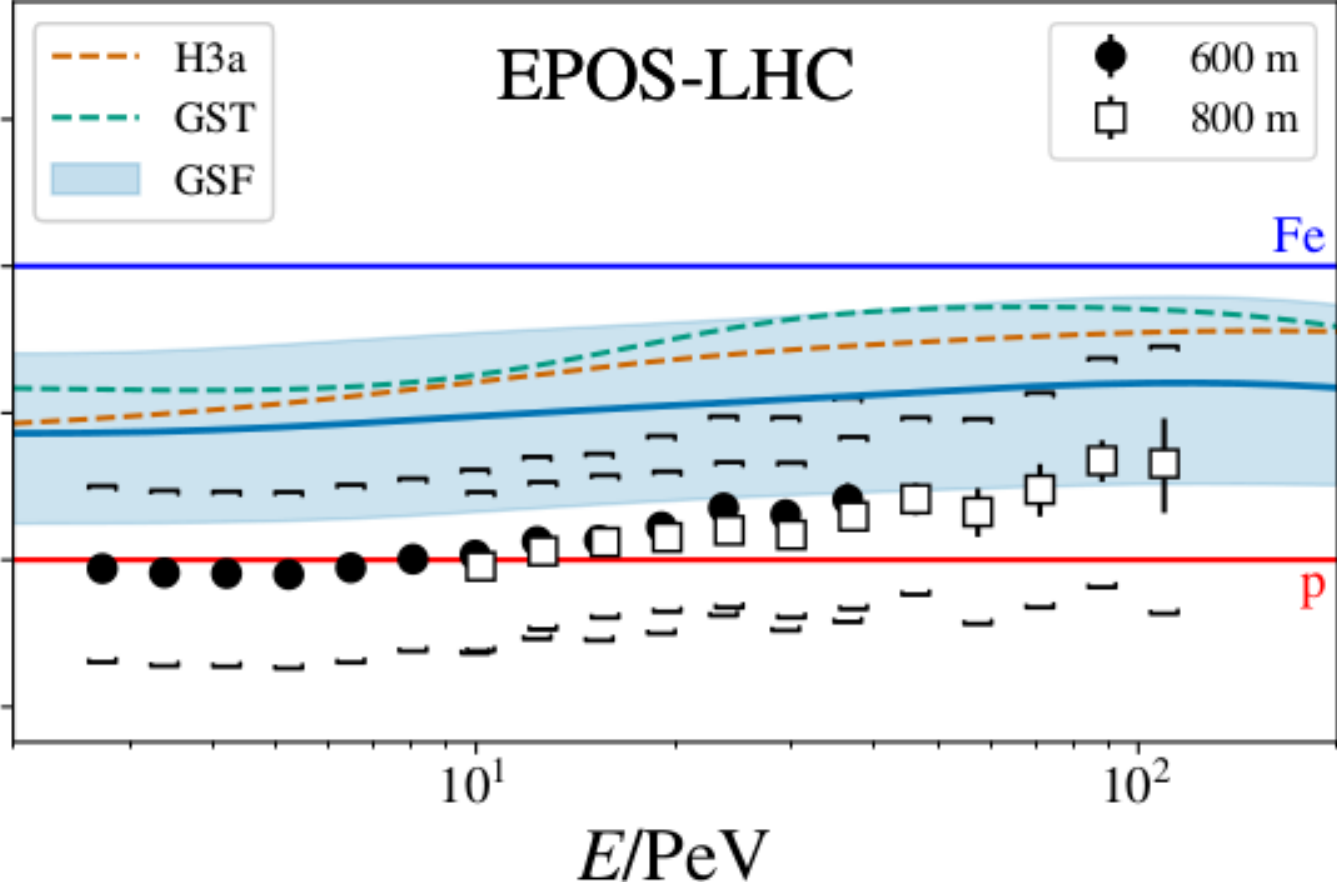}\hspace{-1mm}
\vspace{-1em}
\includegraphics[width=0.34\textwidth]{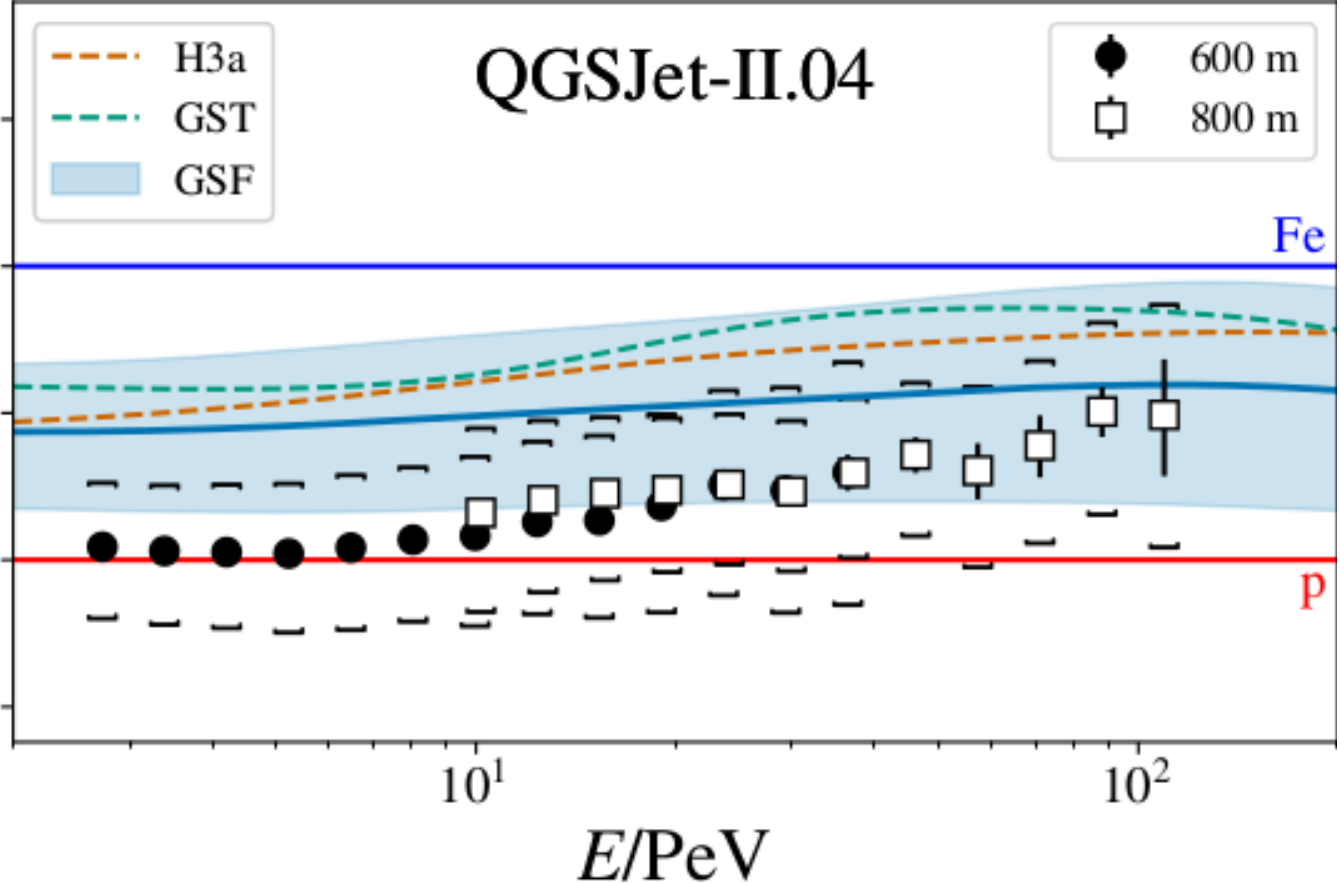}}

\vspace{-0.5em}
\caption{Muon densities in terms of the z-values defined in \Cref{eq:zvalues}, compared to predictions from the hadronic interaction models Sibyll~2.1, EPOS-LHC, and QGSJet-II.04. Also shown are the expectations from the cosmic ray flux models H3a~\cite{Gaisser:2011cc}, GST~\cite{Stanev:2014mla}, and GSF~\cite{Dembinski:2015xtn}. Figures are taken from Ref.~\cite{IceCube:2022yap}.}
\label{fig:RhoMu}
\end{figure}

The muon densities were determined at radial distances of $600\,\mathrm{m}$ for shower energies from $1\,\mathrm{PeV}$ to $40\,\mathrm{PeV}$ and at $800\,\mathrm{m}$ for energies between $9\,\mathrm{PeV}$ and $120\,\mathrm{PeV}$, respectively, using IceTop data taken between June 2010 and May 2013. In order to compare the measured distributions to model predictions, the \emph{z-value} is used which is defined as
\begin{equation}
z=\frac{\log(\rho_\mu)-\log(\rho_{\mu,\mathrm{p}})}{\log(\rho_{\mu,\mathrm{Fe}})-\log(\rho_{\mu,\mathrm{p}})}
\label{eq:zvalues}
\end{equation}
where $\rho_\mu$ is the experimentally measured muon density, and $\rho_{\mu,\mathrm{p}}$ and $\rho_{\mu,\mathrm{Fe}}$ are the muon densities obtained from simulations, assuming a pure proton and iron flux. This is done using CORSIKA~\cite{corsika} with the hadronic interaction models Sibyll~2.1~\cite{Ahn:2009wx}, EPOS-LHC~\cite{Pierog:2013ria}, and QGSJet-II.04~\cite{Ostapchenko:2013pia}. The resulting distributions are shown in \Cref{fig:RhoMu}, compared to the expectations from the cosmic ray flux models H3a~\cite{Gaisser:2011cc}, GST~\cite{Stanev:2014mla}, and GSF~\cite{Dembinski:2015xtn}. While the data are bracketed by the pure proton and pure iron predictions, predictions using the post-LHC models QGSJet II-04 and EPOS-LHC show a larger muon density, requiring a very light mass composition which is in tension with other current experimental results. 

%Measurements from the Pierre Auger Collaboration in Refs.~\cite{Aab:2014pza,Aab:2016hkv,Aab:2020frk}.

\subsection{High-Energy Muons}
\label{sec:HEmuons}

\begin{figure}[tb]
\centering
\vspace{-0.5em}
\includegraphics[width=0.48\textwidth]{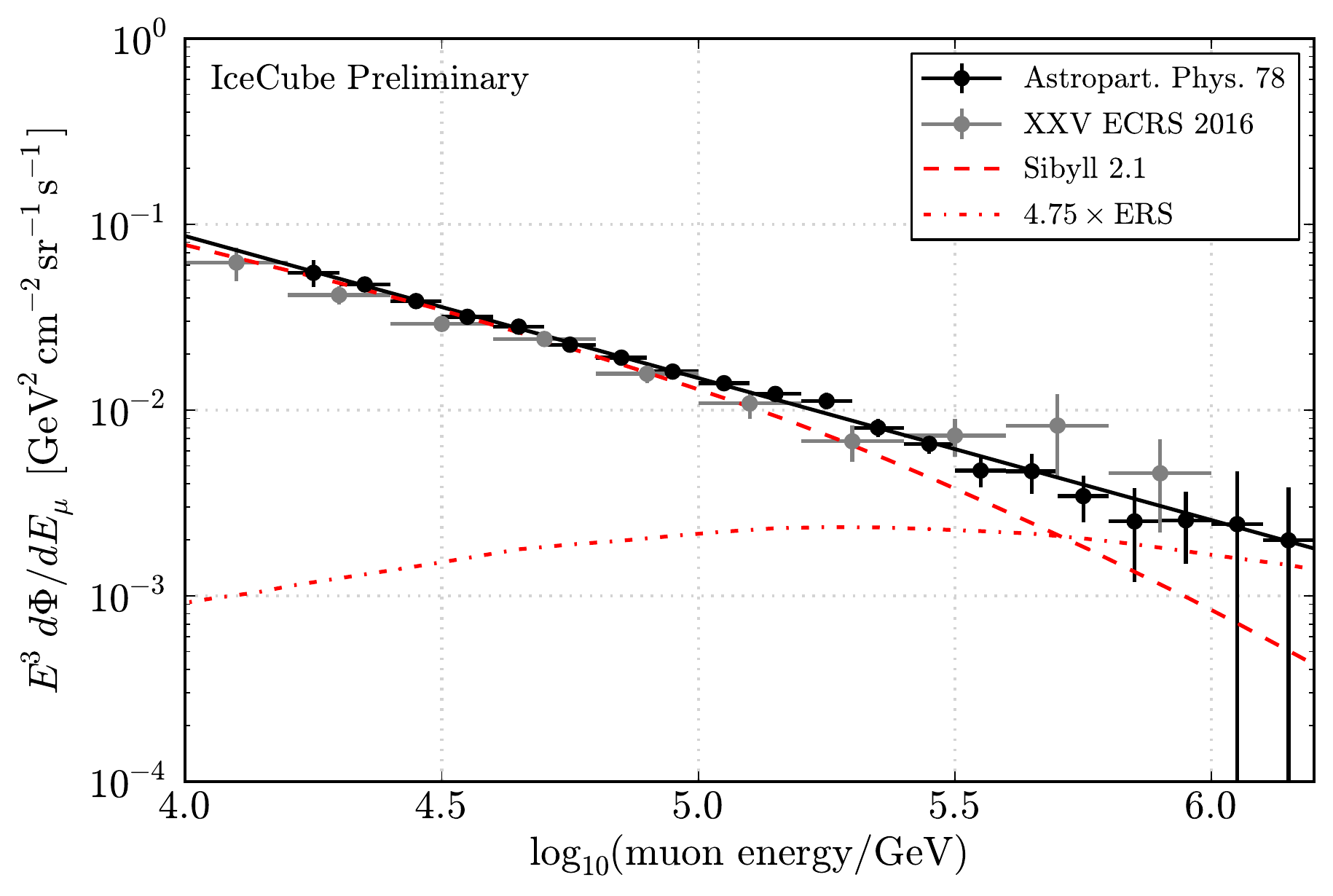}
\includegraphics[width=0.492\textwidth]{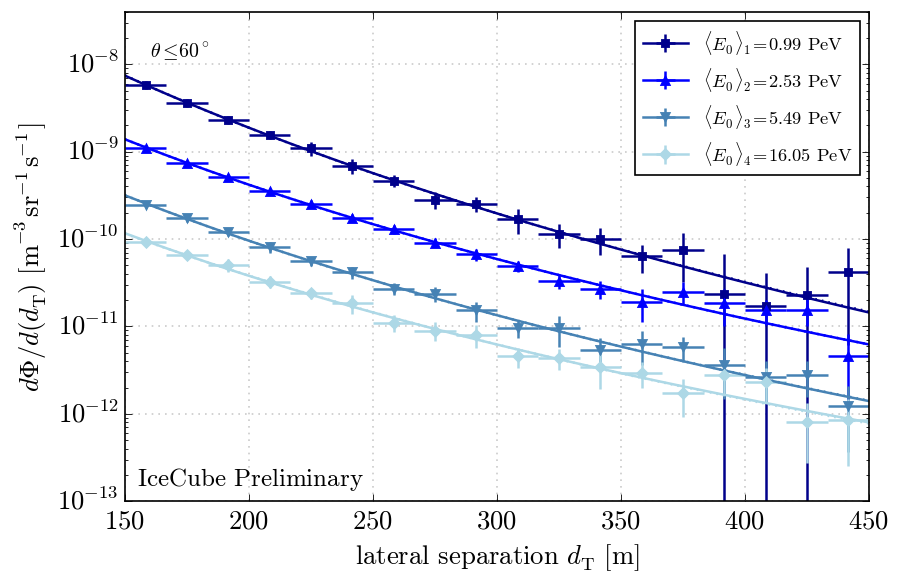}
\caption{\emph{Left:} High-energy muon spectrum measured with IceCube in two separate analyses~\cite{Aartsen:2015nss,Fuchs:2017nuo}. Also shown is the prediction from CORSIKA simulations with Sibyll~2.1 and the best fit prompt flux estimate in terms of the ERS flux~\cite{Enberg:2008te}, assuming H3a as primary cosmic ray flux. \emph{Right:} Lateral distribution of muons with energies above $460 \,\mathrm{GeV}$ for four mean primary cosmic ray energies. Figures from Ref.~\cite{Soldin:2018vak}.}
\label{fig:HEMuons}
\end{figure}

In addition to the measurement of low-energy muons in the IceTop tanks, IceCube is able to measure the high-energy muon content in EAS in the deep-ice detector with high precision. For example, the high energy muon spectrum above $\sim 10\,\mathrm{TeV}$ can be determined using sophisticated energy reconstruction methods based on the energy loss profile in the ice, as described in Refs.~\cite{Aartsen:2015nss,Fuchs:2017nuo}. The resulting muon spectrum was obtained in two separate analyses with the results shown in \Cref{fig:HEMuons} (left) which are compared to the muon flux prediction from CORSIKA using Sibyll~2.1 as the hadronic model. The excess at the highest energies is due to the contribution of prompt muons that are produced in the prompt decay of heavy hadrons and unflavored vector mesons that are not accounted for in Sibyll~2.1. In order to get an estimate of the prompt muon flux, the excess is fit assuming the ERS flux parametrization, described in Ref.~\cite{Enberg:2008te}. Depending on the underlying cosmic ray flux model, the resulting prompt flux estimate is between $\sim 2$ and $\sim 6$ times the ERS flux and a non-existent prompt flux can be excluded with a significance of $\sim 2\sigma$ to $\sim 5\sigma$. However, significant discrepancies in the zenith angle distribution of the high-energy muon flux are observed which are not yet understood and require further investigation (see Refs.~\cite{Aartsen:2015nss,Soldin:2018vak} for details).

In addition, the distribution of laterally separated muons with energies above $460 \,\mathrm{GeV}$ can be determined using IceCube's in-ice detector up to zenith angles of $60^\circ$. These muons are produced from the decay of mesons with large transverse momentum and they can reach separations from the EAS central axis of several $100\,\mathrm{m}$. The muon lateral distribution was derived based on IceCube data taken between May 2012 and May 2014 for four different cosmic ray energy regions using a dedicated reconstruction algorithm, as described in Refs.~\cite{Abbasi:2012kza,Soldin:2018vak}. The resulting lateral distributions are shown in \Cref{fig:HEMuons} (right)~\cite{Soldin:2018vak}. Since this analysis does not use any IceTop information to increase the event statistics, the cosmic ray energy is obtained from the in-ice signals with a resolution of about $0.5$ in $\log_{10}(E/\mathrm{GeV})$. These results agree well with previous measurements by IceCube~\cite{Abbasi:2012kza}, however, discrepancies in the zenith angle distribution are also observed which requires further studies.

\newpage

\subsection{Tests of Hadronic Interaction Models}
\label{sec:coincidentmuons}

\begin{figure}[b]
\centering
\vspace{-1em}
\mbox{\hspace{-2.5em}\includegraphics[width=0.392\textwidth]{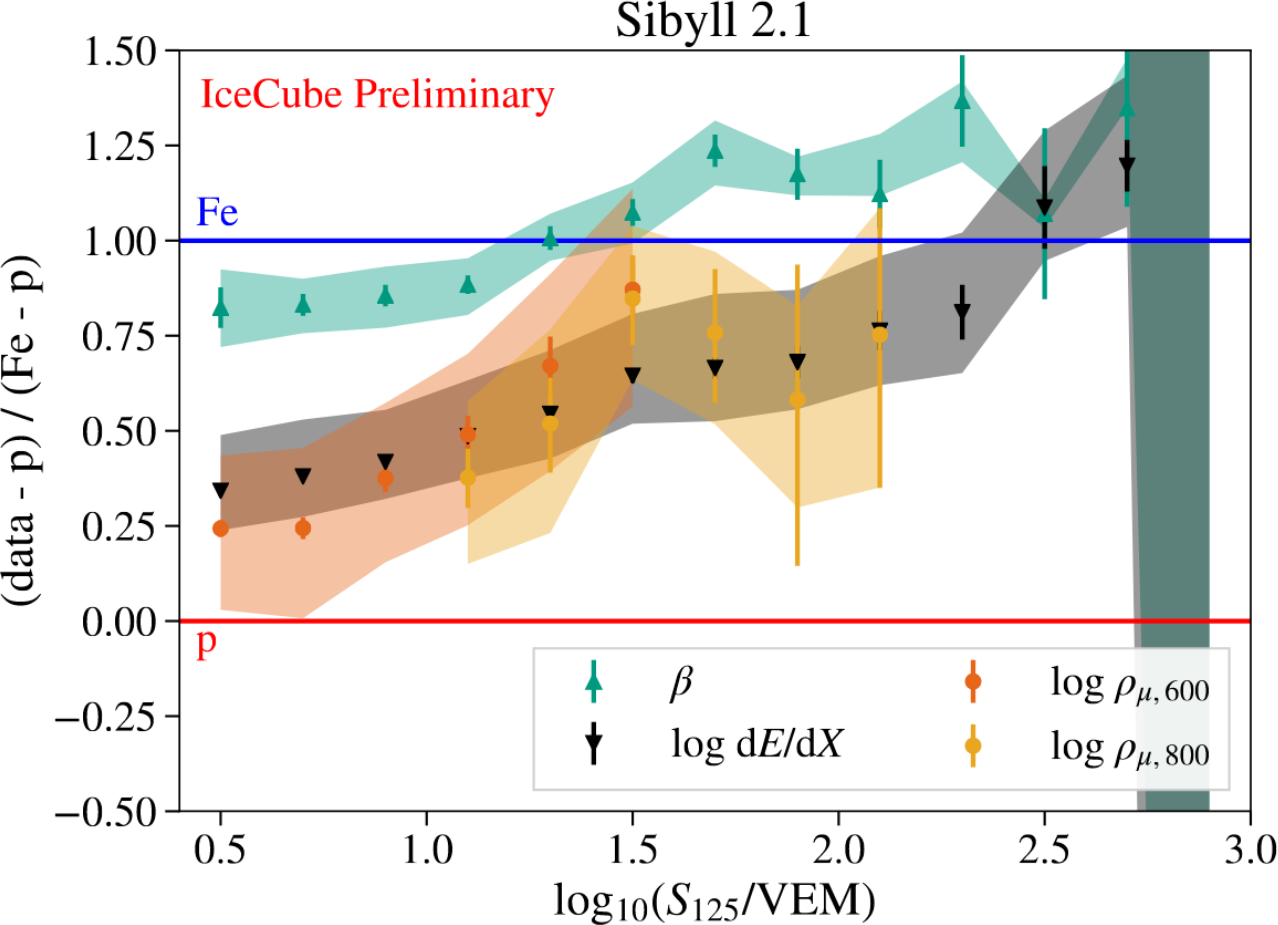}\hspace{-1mm}
\includegraphics[width=0.34\textwidth]{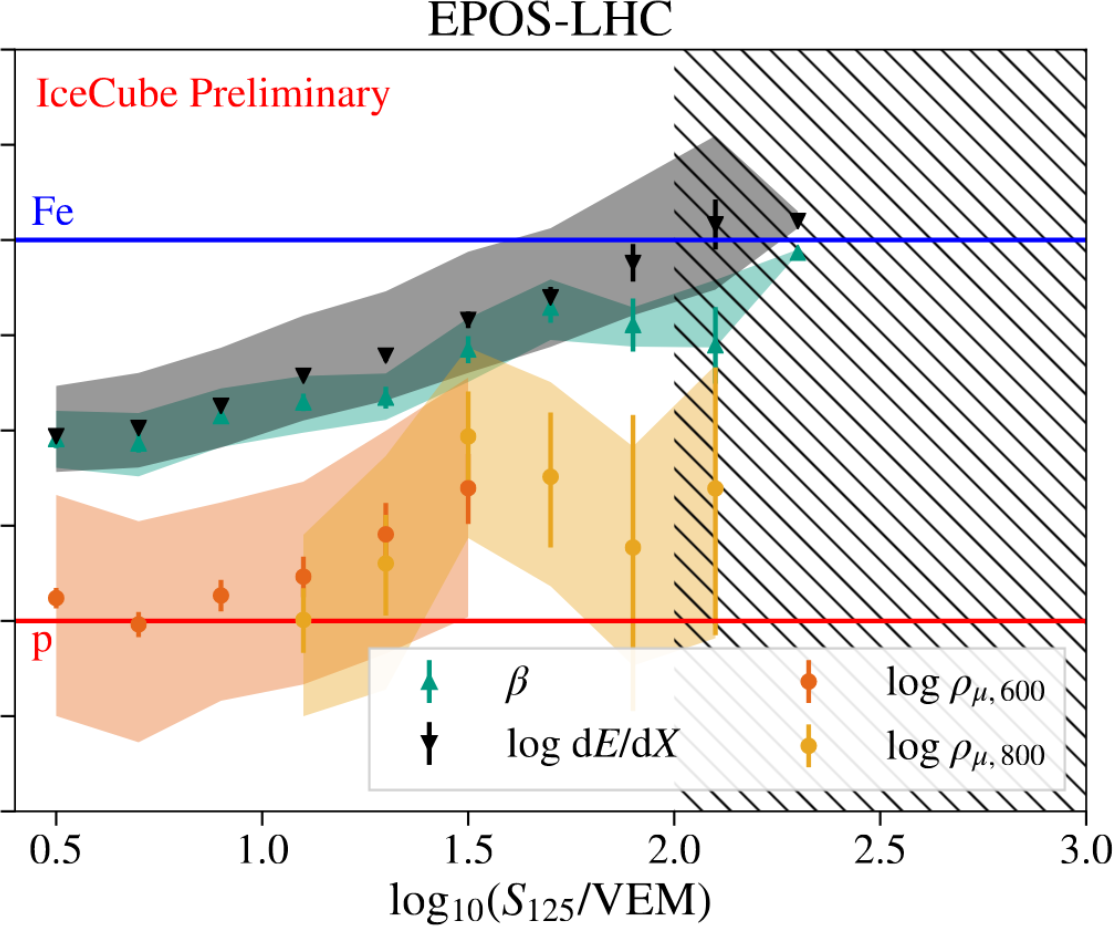}\hspace{-1mm}
%\vspace{-0.2em}
\includegraphics[width=0.34\textwidth]{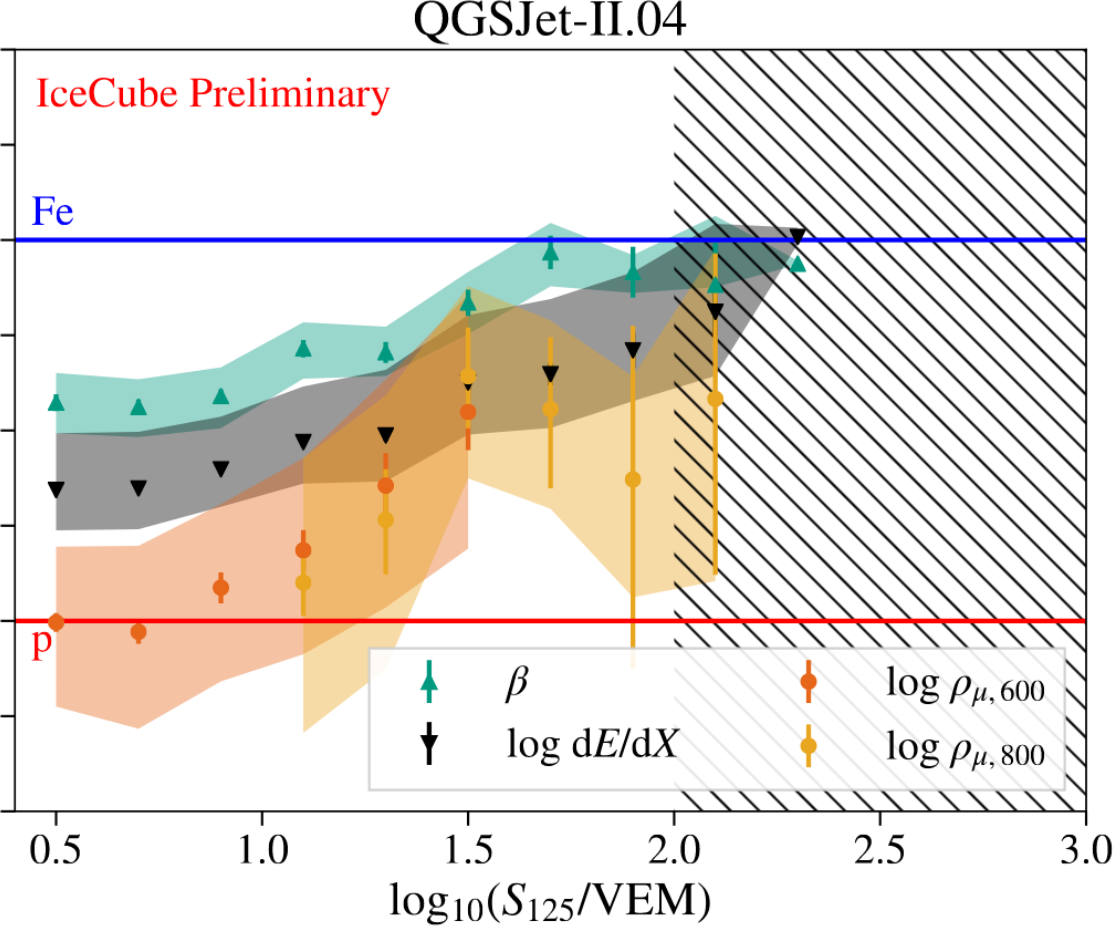}}

\vspace{-0.5em}
\caption{Muon content in air showers in terms of the z-values defined in \Cref{eq:zvalues}, compared to predictions from the hadronic interaction models Sibyll~2.1, EPOS-LHC, and QGSJet-II.04, the latter two with limited statistics. Figures from Ref.~\cite{IceCube:2021ixw}.}
\label{fig:HadrModels}
\end{figure}

Using data from IceCube, various muon sensitive observables can be compared with predictions from simulations to check their consistencies. From $10\%$ of data taken between May 2012 and May 2013 coincident events were selected that have signals in both IceTop and the in ice detector~\cite{IceCube:2021ixw}. For this event selection the muon densities are determined as described in \Cref{sec:GeVmuons}. In addition, the slope parameter, $\beta$, and the deposited energy, $dE/dX$, along the reconstructed trajectory in the ice at a slant depth of $1500\,\mathrm{m}$ are obtained as described in \Cref{sec:spectrum} and \Cref{sec:composition}. These parameters are compared to CORSIKA simulations of proton and iron showers using the z-values defined in \Cref{eq:zvalues}. The resulting distributions as a function of $\log_{10}(S_{125}/\mathrm{VEM})$ are shown in \Cref{fig:HadrModels} for the hadronic interaction models Sibyll~2.1~\cite{Ahn:2009wx}, EPOS-LHC~\cite{Pierog:2013ria}, and QGSJet-II.04~\cite{Ostapchenko:2013pia}. For all models the z-values increase with increasing $S_{125}$, consistent with a cosmic ray mass that becomes heavier towards high energies. Although the general behavior agrees with the results discussed in \Cref{sec:composition} there are significant inconsistencies between the different observables. The differences of the muon densities between the models are qualitatively consistent with the results shown in \Cref{fig:RhoMu} where the post-LHC models EPOS-LHC and QGSJet-II.04 require a very light mass composition. The high-energy muon content measured by $dE/dX$, however, is consistent between the models within the uncertainties. Further studies of the low- and high-energy muon content in EASs with IceCube are currently ongoing in order to provide additional tests and constraints for hadronic interaction models in the future.%Test \cite{Soldin:2021wyv} 

%WHISP \cite{Dembinski:2019uta,Cazon:2020zhx,Soldin:2021wyv} and the \emph{Muon Puzzle} \cite{Albrecht:2021yla}.

\newpage

\section{Conclusions}

In this report, recent results from measurements of the cosmic ray spectrum between $250\,\mathrm{TeV}$ to $1\,\mathrm{EeV}$ in IceCube were presented and an analysis of the mass spectrum above $1\,\mathrm{PeV}$ was shown. Moreover, IceCube provides unique opportunities to measure the muon content in EASs. A measurement of the densities of muons with energies around $1\,\mathrm{GeV}$ in IceTop was presented, as well as measurements of high-energy muons in IceCube's deep ice detector. These measurements were used to test predictions from hadronic interaction models, considering various cosmic ray flux models. While the predictions of the low-energy muons in EASs differ between models, the high-energy muon content agrees within uncertainties. 

In the future, new scintillator and radio detectors at the surface will further improve the capabilities for cosmic ray measurements in IceCube~\cite{IceCube:2021ydy}. In addition, the construction of a new surface array is planned in the context of IceCube-Gen2~\cite{IceCube-Gen2:2021aek}. These detectors will increase the energy resolution of IceCube and improve the separation of the electromagnetic and muonic components of EASs. In turn, this will reduce uncertainties in the measurement of the cosmic ray mass composition and improve studies of the muon content in EASs.

%\newpage

\section*{Acknowledgements}
%I would like to thank my colleagues in the IceCube Collaboration for discussions and support.

%\vspace{-1em}

% TODO: include author contributions
%\paragraph{Author contributions}
%This is optional. If desired, contributions should be succinctly described in a single short paragraph, using author initials.

% TODO: include funding information
\paragraph{Funding information}
D.S. acknowledges the support from the US NSF Grant PHY-1913607.

\bibliography{references.bib}

\nolinenumbers

\end{document}